\documentclass[11pt]{article}

\usepackage[utf8]{inputenc}
\usepackage[T1]{fontenc}
\usepackage{lmodern}
\usepackage{microtype}

\usepackage{longtable,booktabs,array,ragged2e}
\usepackage{calc}
\usepackage{caption}
\usepackage{float}
\usepackage{enumitem}

\usepackage{graphicx}
\usepackage{xcolor}

\usepackage{amsmath,amssymb}

\usepackage{bookmark}
\usepackage{hyperref}
\usepackage{xurl}

\hypersetup{
  pdftitle={Developing the Criteria for Credible AI-assisted Systems: The Case of Mapping and Lifecycle Modeling},
  pdfauthor={Shaena Ulissi, Andrew Dumit, P. James Joyce, Krishna Rao, Steven Watson, Sangwon Suh},
  hidelinks,
  colorlinks=false,
  pdfcreator={LaTeX}
}

\title{Criteria for Credible AI-assisted Carbon Footprinting Systems: The Cases of Mapping and Lifecycle Modeling}

\author{
  Shaena Ulissi\textsuperscript{1*} \and
  Andrew Dumit\textsuperscript{1} \and  
  P. James Joyce\textsuperscript{1} \and
  Krishna Rao\textsuperscript{1} \and
  Steven Watson\textsuperscript{1} \and
  Sangwon Suh\textsuperscript{1*}
}

\usepackage[margin=1in]{geometry}
\renewenvironment{abstract}{\small\quotation\textbf{Abstract}\par}{\endquotation}

\begin{document}
\nocite{*}
\maketitle

\begin{center}
\small
\textsuperscript{1}Watershed Technology Inc\\
\textsuperscript{*}Corresponding authors: \texttt{shaena@watershedclimate.com}, \texttt{sangwon@watershedclimate.com}

\end{center}

\begin{abstract}
As organizations face increasing pressure to understand their corporate and products' carbon footprints, artificial intelligence (AI)-assisted calculation systems for footprinting are proliferating, but with widely varying levels of rigor and transparency. Standards and guidance have not kept pace with the technology; evaluation datasets are nascent; and statistical approaches to uncertainty analysis are not yet practical to apply to scaled systems. 
We present a set of criteria to validate AI-assisted systems that calculate greenhouse gas (GHG) emissions for products and materials. We implement a three-step approach: (1) Identification of needs and constraints, (2) Draft criteria development and (3) Refinements through pilots. The process identifies the needs for distinguishing three use cases of AI applications: Case 1 focuses on AI-assisted mapping to existing datasets for corporate GHG accounting and product hotspotting, automating repetitive manual tasks while maintaining mapping quality. Case 2 addresses AI systems that generate complete product models for corporate decision-making, which require comprehensive validation of both component tasks and end-to-end performance. We discuss the outlook for Case 3 applications, systems that generate standards-compliant models, and acknowledge the challenges and approaches that remain to be developed. We find that credible AI systems can be built and that they should be validated using system-level evaluations rather than line-item review,  with metrics such as benchmark performance, indications of data quality and uncertainty, and transparent documentation.
This approach may be used as a foundation for practitioners, auditors, and standards bodies to evaluate AI-assisted environmental assessment tools. By establishing evaluation criteria that balance scalability with credibility requirements, our approach contributes to the field's efforts to develop appropriate standards for AI-assisted carbon footprinting systems. 

\end{abstract}

\section{Introduction}
\label{sec:introduction}

The past few years have seen explosive growth in AI applications for carbon accounting, with published studies spanning machine learning (ML) methods to fill gaps in life cycle inventories \cite{cullen_machine_2024,zhao_data-centric_2025}, environmental impact factor matching for life cycle assessment (LCA) \cite{balaji_emission_2025,balaji_flamingo_2023,castle_entity_2025}, systematic and automated selection of LCA dataset for the automotive industry \cite{sturm_systematic_2025}, autonomous sustainability assessment with agents \cite{zhang_towards_2025}, and automated financial mappings dataset development \cite{dumit_atlas_2024,balaji_emission_2025}, among others. 

Concurrent with this research activity, specialized workshops and conference proceedings have proliferated across major venues -- from the AI4LCA Workshop \cite{ai4lca_2024} to dedicated sessions at International Society for Industrial Ecology (ISIE) \cite{isie_2025}, Forum for Sustainability through Life Cycle Innovation (LCIC) \cite{lcic_workshop_2024}, and Neural Information Processing Systems (NeurIPS) \cite{neurips_tackling_2024,neurips_tackling_2025}. These demonstrate widespread academic and industry recognition of AI's transformative potential for environmental assessment. 

This rapid proliferation has created an urgent quality control challenge. An increasing number of consulting firms and software providers are deploying AI-enabled product carbon footprinting (PCF) tools with widely varying methodological approaches and quality standards. Traditional verification methods---designed for human-led, transparent LCA processes---are proving inadequate for AI systems that make thousands of algorithmic decisions within opaque model architectures. The result is a growing ecosystem of AI-generated carbon footprints with no systematic way to assess their reliability, comparability, or fitness for different use cases.

Despite all the recent technical studies, no existing studies that we are aware of propose consistent and feasible evaluation criteria that carbon footprinting providers can use when implementing and verifying AI-based systems. Current literature focuses primarily on model development and performance optimization, leaving a critical gap in establishing industry-wide standards for quality assessment, documentation requirements, and verification protocols. Our approach addresses that gap by providing an initial set of evaluation criteria designed specifically for the unique challenges of AI-enabled mapping and lifecycle modeling.

This paper describes our approach to developing evaluation criteria for AI-assisted carbon footprinting, informed by ongoing research and validation of an AI system for PCF generation. We present our methodology for establishing evaluation criteria as a contribution to the field's efforts to develop appropriate standards for AI-assisted environmental assessment tools. 

\section{Methods}
\label{methods}
We developed our evaluation criteria for AI-assisted carbon footprinting (AI-CF) systems through a systematic, multi-step process. Our approach combined (1) needs and constraints identification through research and expert interviews, (2) design and development of draft criteria, and (3) refinement through iterative pilot testing. This methodology intentionally avoids specifying particular model architectures, focusing instead on outcome-based criteria that can accommodate the rapid evolution of AI technologies. Our approach recognizes that different applications require distinct use cases of scrutiny and documentation. We focus on criteria for two primary use cases - Case 1 and Case 2 - that represent the current state of feasible AI-assisted carbon footprinting, and introduce a third use case. We provide detailed evaluation criteria for both levels and discuss the outlook for alternative applications. 

\subsection{Identification of needs and constraints}
\label{methods-1}
The first phase of our work involved preparing the groundwork for the criteria. We began with a review of existing literature, standards, and datasets. Our goal was to identify established best practices and understand the current state of the science. We reviewed standard protocols such as the International Standards Organization (ISO) 14044, the GHG Protocol Product Standard, and the World Business Council for Sustainable Development (WBCSD) PACT framework to understand typical requirements for carbon accounting. For data, we sought reputable sources that could represent a ground truth for our evaluations, including validated financial mappings and validated PCFs. We also reviewed the literature on topics such as uncertainty analysis and Life Cycle Inventory (LCI) data modeling to understand the field's aspirational goals. We also reviewed recent works on AI for PCFs such as those cited in the Introduction.

Simultaneously, we conducted an expert elicitation process to gain practical insights from the industry. This consultation began over the period from June 2024 and is ongoing as of August 2025. Experts were chosen based on their records of expertise in the relevant fields, which were environmental verification, life cycle assessments, and AI model use. For perspectives on verification, we consulted via video conferencing meetings and written communications with two professional verification firms with decades of experience in environmental audits covering two major geographies (Europe and North America, respectively). We also reviewed articles from to understand initial perspectives on AI-assisted verification\cite{kpmg_2025}. In parallel, we elicited feedback from over a dozen corporate practitioners actively engaged in carbon accounting to understand what would build their trust in an AI system. These practitioners included LCA experts and procurement experts from large enterprises that span multiple industries, including automotive manufacturing, sporting goods, energy generation, agricultural commodities,packaging, consumer goods, and other manufacturing. This process was iterative, involving multiple in-person workshops, video-conference discussions, and product-feedback use with these companies as we developed our own AI-CF systems. Throughout, we worked with AI domain experts within our company and via more informal conversations at conferences including NeurIPS\cite{neurips_2024} and the AI Engineers World Fair\cite{aie_2025} to understand AI practices outside the domain of sustainability. These conversations with both domain and technical experts were crucial in shaping our approach to focus on outcomes rather than specific technical implementations. 

\subsection{Design and Development of Draft Criteria}
\label{methods-2}
Following the foundational phase, it became clear that different application types present different needs for evaluation metrics and documentation. We synthesized our findings to develop a preliminary set of criteria for two distinct use cases and identified a third use case. We framed these as Case 1: AI-Assisted Mapping and Case 2: Automated Modeling. Case 3: Standards compliant modeling was also identified as a common use case, but standardized criteria were not feasible to develop. While Case 1 systems focus on improving the efficiency and consistency of mapping to a single activity, Case 2 systems use AI to generate more complete and accurate product models. These use cases are based on the following key learnings about domain challenges. 

Conventional approaches that many existing practitioners and software providers use for mapping products and materials to GHG emission factors create systematic barriers to effective carbon management. These approaches undermine companies' ability to make informed procurement decisions, track progress over time, and defend their carbon accounting choices to stakeholders. These limitations become particularly problematic as companies scale their carbon footprinting efforts and face increasing scrutiny from regulators, investors, and customers who expect consistent, defensible methodologies. Industry-average, spend-based emission factors are typically the starting point for corporate supply chain carbon management. Mapping to these factors requires knowledge of the boundaries of the factors and consistent application across often many thousands of procurement inputs. 

Human expert mapping to detailed secondary datasets addresses the granularity problem but creates new scalability and consistency challenges that limit practical implementation. A widely used activity-based data source -- ecoinvent \cite{ecoinvent}  -- has upwards of 2,000 unique reference products. Time constraints make comprehensive mapping prohibitively expensive---with expert practitioners requiring 10+ minutes per material row, companies with large portfolios (often 10,000+ unique materials) \cite{wakelin_how_2021} face high mapping costs. Inconsistency across practitioners means that different experts evaluating identical materials can take substantially different approaches (e.g., we have found this with internal expert reviews, and the conclusion is supported by studies such as \cite{kuczenski_prototypes_2021} for full LCA setup and \cite{konradsen_same_2024} for environmental product declarations [EPDs]). This creates systematic bias in carbon footprint comparisons between products, suppliers, or time periods. This undermines the repeatability that auditors and standards require. These inconsistencies become amplified when companies scale their programs and involve multiple team members or external consultants, making it difficult to defend methodological choices or ensure comparable results across their portfolio. 

Beyond AI-assisted mapping, there is a need for additional capabilities. The scaling up of PCF preparation is already underway, driven by both regulatory pressures and industry recognition of the limitations of current proxy-based approaches. Companies want access to detailed modeling for process design and engineering and procurement decisions. This can be inferred from recent articles from companies that discuss their procurement and materials strategies as related to PCFs \cite{trek_2025,basf_2025,unilever_2022,rivian_2025}, and was confirmed through our pilot discussions with other companies. Given the resources required to prepare PCFs conventionally is unscalable, AI-assisted PCF products are promoted as potential solutions. Amazon \cite{wang_2025} and Meta \cite{rivalin_2024}both have recent case studies using AI to assist in BOM breakdowns and mappings.   

We also identified the need for a Case 3 set of criteria and capabilities to meet stated standards, frameworks, and methodologies such as ISO 14040/14044\cite{iso_14044}, 14067\cite{iso_14067}, and the GHG Protocol Product Standard\cite{ghgp_product}. Regulatory initiatives are creating mandatory requirements: for example, the European Union (EU)'s Corporate Sustainability Reporting Directive and the California Corporate Greenhouse Gas Reporting Program will require detailed Scope 3 emissions data, while emerging product-level carbon labeling schemes demand standardized carbon footprint calculations. Industry coalitions are building the infrastructure to support this transition---the World Business Council for Sustainable Development (WBCSD) Partnership for Carbon Transparency (PACT) framework specifically aims to enable suppliers of all sizes, across all supply chain tiers, to calculate and share product level carbon footprint data with minimal barriers \cite{wbcsd_pact}. However, as a 2025 Organization for Economic Cooperation and Development (OECD) analysis of agri-food supply chains demonstrates, significant methodological and data quality challenges remain as these systems need to scale rapidly \cite{oecd}. 

Identification of precise evaluation criteria for Case 3 was not possible because each standard contains its own unique set of requirements. While many of the criteria and capabilities identified in Case 2 may not be needed for standards compliance, different efforts might be needed to ensure an AI-CF complies with specified reporting requirements. For example, a PACT-aligned model would need to track the percent of primary data. Certain standards require the use of specific numeric assumptions or lifecycle stages. Rules range from generic (e.g., ISO 14044, ISO 14067, GHG Protocol Product Standard) to sector-specific (e.g., Catena-X for automotive parts \cite{catena}, EPEAT for low-carbon solar eco-labeling \cite{epeat}, the International Dairy Federation for dairy \cite{idf}, six distinct methods for packaging \cite{tascione_comparative_2024}, among many others). These requirements vary by industry and use case, and there would likely be a human in the loop somewhere along the process to define to the AI system what requirements to follow.

\subsection{Refinement through iterative pilot testing}
\label{methods-3}
The final phase involved an iterative process of pilot testing and refinement. We applied many potential criteria to our own AI-CF systems and used the results to fine-tune our framework. We tested the application with the same practitioners and companies from which we elicited feedback in the first step, and added additional companies that covered other industries including pharmaceuticals, apparel, retail, food and beverage, and chemicals. 

This process led us to reject several initial ideas that were not yet feasible or mature. For instance, we chose not to propose specific quantitative targets for performance across all systems because ground truth benchmark datasets are not yet widely available. Similarly, we found that requiring quantitative uncertainty requirements substantially exceeded the capabilities of current non-AI state-of-the-art methods. The validated PCFs we reviewed did not reach the bar of state of the science for methods such as uncertainty, but it is helpful to know where the field hopes to move. We also identified cases where criteria were insufficient on their own; for example, benchmarking end-to-end carbon footprints was insufficient without supplemental criteria to ensure a correct result wasn't achieved for the wrong reasons. Our final set of criteria is a direct result of this refinement process. The resulting criteria, along with the detailed performance metrics and validation requirements for both Case 1 and Case 2 systems, are presented in the following sections.

\section{Results}
\label{sec:results}
Results are described as follows, along with a more in-depth discussion of the final criteria. Three converging forces that we found through the foundational research in Step 1 paint a clear need for standardized evaluation criteria: First, regulatory frameworks requiring detailed product-level emissions data will necessitate verification standards that can handle AI-generated outputs at scale. Second, the carbon accounting industry's existing quality assurance infrastructure---built around manual verification and peer review---cannot economically scale to meet the volume of AI-generated footprints entering the market. Third, as AI tools become standard practice rather than experimental applications, practitioners and verifiers need clear benchmarks to distinguish reliable systems from unreliable ones, particularly as these outputs increasingly influence major procurement and investment decisions. 

The AI domain expert discussions led us to conclude that AI models continue to evolve rapidly, and our approach intentionally avoids specifying particular model architectures or training approaches. Instead, we focus on outcome-based evaluation criteria that can accommodate different technical implementations. While current large language models cannot independently generate validated PCFs, neither do AI-assisted systems necessarily require fully custom model development to achieve useful domain expertise, though that has recently been attempted. \cite{li_pcf-rwkv_2025, martinez-ramon_frameworks_2024} 

For Case 2, we identified several example capabilities beyond mapping that would be necessary to build trust in an automated system, including the ability to:
\begin{itemize}
    \item Deconstruct a product into constituent materials: A secondary dataset often does not have an exact match for a given product or material. AI can be used to generate a synthetic bill of materials (BOM).
    \item Estimate missing flows: Reasonably estimate process, energy, and transportation flows (and yield rates).
    \item Support continuous learning: Improve analysis quality over time as more data becomes available, with the system becoming increasingly valuable as it processes more information. This is in contrast with static PCFs.
\end{itemize}

\begin{figure}[H]
    \centering
    \includegraphics[width=0.75\linewidth]{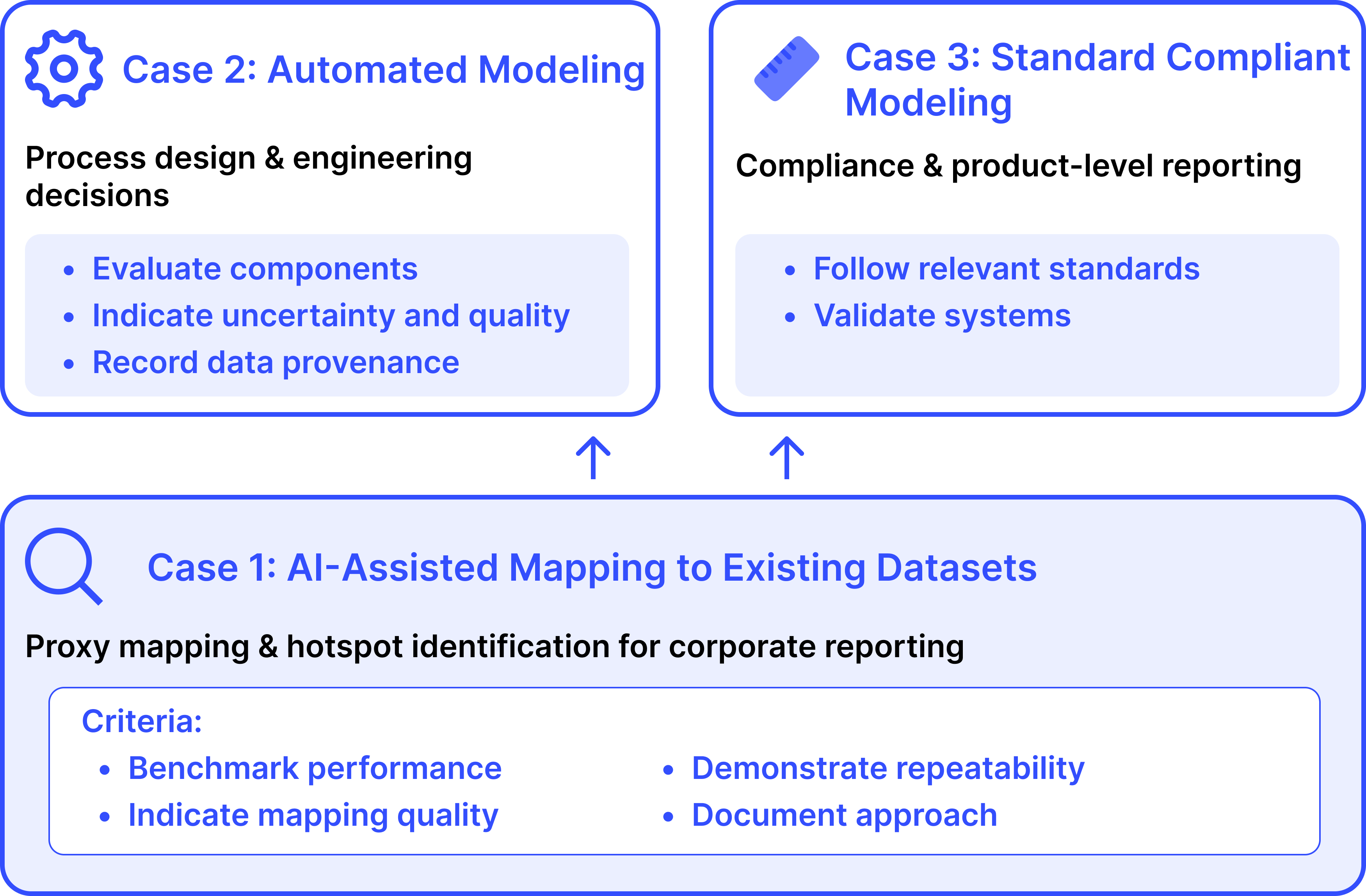}
    \caption{Overview of AI-CF use cases and criteria}
    \label{fig:use_case_framework}
\end{figure}

Figure~\ref{fig:use_case_framework} provides an overview of the criteria and use cases that are described further below. For Case 1 and Case 2, we provide performance and robustness criteria as well as transparency and traceability criteria and validation requirements that effective AI-CF systems should demonstrate.  Criteria are described as system-level or material-level depending on whether they shall be applied to the AI system as a whole as a one-time exercise (that is updated upon all major system updates) or applied to each carbon footprint calculation. 

\subsection{Case 1 Criteria}
\label{results-1}
Effective AI-assisted mapping systems should demonstrate the criteria in Tables~\ref{tab:case1_performance} and~\ref{tab:case1_transparency}. Additional details on criteria are presented below the tables. 

\begin{table}[H]
\centering
\small
\caption{Case 1 mapping performance and robustness criteria}
\label{tab:case1_performance}
\begin{tabular}{|p{.5cm}|p{2cm}|p{4cm}|p{8cm}|}
\hline
\textbf{\#} & \textbf{Application} & \textbf{Name} & \textbf{Summary} \\
\hline
1.1 & System & Benchmark performance & AI-generated mappings should demonstrate their ability to map to defensible choices relative to expert mappings when tested against a validation set of products with established "ground truth" mappings. \\
\hline
1.2 & System & Consistency across material types and input formats & Mapping performance should remain stable across different material industry categories and levels of complexity. If it is not stable, this should be disclosed and caveated appropriately if the model is used for materials that are outside its domain expertise. \\
\hline
1.3 & System & Appropriate granularity and proxy approach & Where exact matches aren't available, the systems should apply appropriate proxy selection logic that aligns with LCA best practices. The system should map to the most specific appropriate activity dataset rather than defaulting to overly broad categories. \\
\hline
1.4 & System & Repeatability & Small variations in product descriptions or specifications should not result in dramatically different mappings. In addition, the same material should generally be mapped consistently. \\
\hline
1.5 & Material & Match quality indication & The system should display to the user the uncertainty level and highlight poor matches between the input material and the mapped proxy dataset; these would be places where, if material, the user should attempt to augment their data input or move to Case 2 carbon footprinting. \\
\hline
\end{tabular}
\end{table}

\textbf{Criteria Details:}

\textbf{[1.1] Benchmark performance:} The precise benchmark result will depend on the complexity of the products and available datasets to map to, so we do not propose a specific target for this metric here -- but if a labeled dataset were to be made available such as \cite{balaji_emission_2025, dumit_atlas_2024}, it could be used to set a common target. For example, mappings where there are exact matches should likely achieve near 100\% benchmark performance because there is high agreement among experts and clear choices. If experts systematically map incorrectly, this benchmark may also be less than 100\%. Complex plastic materials or chemicals mapped to a data source like ecoinvent should also achieve a high benchmark performance, but a result less than 100\% is likely acceptable as there will be less agreement among experts as to the best and defensible mappings. A system with a lower performance may be acceptable if automated mappings are only used for small parts that are not material to a given company's purchased materials. To demonstrate this performance, the model provider should prepare documentation, visuals, or calculations that evaluate the performance at this mapping step. The performance evaluations should be repeated whenever the model undergoes a major update, to ensure there are no regressions in quality. 

\textbf{[1.2] Consistency across material types and input formats:} A distribution of benchmark performance by industry classification may be used to demonstrate this consistency. Additional considerations that arise from actual company material data may be used to generate additional benchmarks and evaluations. For example, company material data is often provided with abbreviations or industry-specific terms and acronyms, in different languages, or with misspellings. 

\textbf{[1.3] Appropriate granularity and proxy approach:} Carbon accounting and materials mapping have no universally accepted playbook; practitioners improvise with ad-hoc heuristics, making results hard to reproduce or audit. By adapting a clear, step-by-step method for turning messy material descriptions into consistent mappings and codifying these rules, the resulting approach can be consistent, explainable, and audit-ready. This process framework and example test cases of how the system follows the framework should be provided to assist verifiers, auditors, or reviewers in understanding the system. 

\textbf{[1.4] Repeatability:} Given the non-deterministic nature of AI/machine learning (ML) methods, this metric will rarely be 100\% consistent, but it should be high enough to be useful for decision makers and at least as consistent as human experts. This can be demonstrated with repeatability evaluations that consider test cases across various material types, where the same material or similar materials are mapped multiple times. Several metrics may be relevant, such as 

\begin{equation}
\text{repeatability score} = \frac{\text{N(most common prediction)}}{\text{N(total predictions)}}
\label{eq:repeatability}
\end{equation}

\begin{equation}
\text{normalized entropy} = \frac{-\sum_{i} p_i \log_2(p_i)}{\log_2(\text{N(unique predictions)})}
\text{ where } p_i = \frac{\text{N(prediction}_i\text{)}}{\text{N(total predictions)}}
\label{eq:normalized_entropy}
\end{equation}

\begin{equation}
\text{N\_unique} = \text{N(unique predictions per material)}
\label{eq:unique_predictions}
\end{equation} 

\textbf{[1.5] Match quality indication:} Many types of uncertainty affect GHG analyses, including estimation uncertainty, parameter uncertainty, statistical uncertainty, model uncertainty, systematic uncertainty, and scientific uncertainty. There are many potential paths to display and calculate uncertainty in LCA e.g. \cite{mendoza_beltran_quantified_2018, qiao_stochastic_2025, cullen_reducing_2024}. To date, many existing LCA datasets have used a Pedigree matrix or data quality indicators (DQIs). Parametric Monte-Carlo propagation, or model-centred approaches are more data-heavy approaches that may result in more statistically significant results. In addition, global sensitivity analysis to quantify which inputs dominate footprint variance may be useful\cite{adams_parameters_2025}. The GHG Protocol Technical Working Groups suggest that uncertainty and data quality will be seen with greater importance in the upcoming standards update\cite{ghgp_2025}. There is an additional key level of uncertainty that arises from the match quality between the dataset and the input material. For proxy mapping to a single dataset, most likely a qualitative uncertainty framework such as DQI would suffice to inform the user of the relative quality of each of their mappings. We found that this last type of uncertainty of match quality is typically overlooked in LCA but can have a strong impact on overall emissions representation. 

AI-assisted mapping systems should provide the indicators or written narratives shown in Table~\ref{tab:case1_transparency} at a system level, in addition to or as part of the metrics described above. 

\begin{table}[H]
\centering
\small
\caption{Case 1 System-level transparency and traceability criteria}
\label{tab:case1_transparency}
\begin{tabular}{|p{.5cm}|p{5cm}|p{9cm}|}
\hline
\textbf{\#} & \textbf{Name} & \textbf{Summary} \\
\hline
1.6 & Mapping methodology documentation & Explanation of how the AI system interprets product characteristics and selects appropriate activity datasets. \\
\hline
1.7 & Data source identification & Documentation of the activity-based datasets used (e.g., ecoinvent version) and any supplementary data sources. \\
\hline
1.8 & Version control & Tracking of mapping algorithm versions and activity dataset versions to ensure reproducibility. \\
\hline
1.9 & Decision logic transparency & Visibility into key decision points and thresholds used in the mapping process. \\
\hline
1.10 & Audit trail & Ability to trace how specific product characteristics led to particular activity dataset selections. Given the non-deterministic character of large language model (LLM)-based systems, this audit trail may be at the bulk level or on example datasets rather than specific to each individual product or material. \\
\hline
1.11 & Continuous improvement & Description of mechanisms for incorporating feedback to improve mapping accuracy over time. \\
\hline
1.12 & Position in GHGP hierarchy & Clear positioning of this approach within the GHG Protocol Scope 3.1 data quality hierarchy, demonstrating improvement over broad sector averages while acknowledging that Cases 2 or 3 are needed to move further up the hierarchy. \\
\hline
\end{tabular}
\end{table}

For Case 1 applications, validation should include: 
\begin{itemize}
    \item \textbf{Expert confirmation of benchmarking and evaluation results from above:} Confirmation from an expert such as a verifier that the system continues to meet the criteria described above, such as assessment of a sample of AI-mapped results against human expert mapping or benchmark data sets to ensure continued accuracy.
    \item \textbf{Confirmation} of the documentation requirements described above.
\end{itemize}

\subsection{Case 2 Criteria}
\label{results-2}

All of the Case 1 criteria for proxy mappings are required as prerequisites to the end-stage mappings in Case 2 models. In addition to those criteria, automated modeling systems should demonstrate the following criteria (Tables~\ref{tab:case2_performance} and~\ref{tab:case2_transparency}). 

\begin{table}[H]
\centering
\small
\caption{Case 2 performance and robustness criteria}
\label{tab:case2_performance}
\begin{tabular}{|p{.5cm}|p{2cm}|p{4cm}|p{8cm}|}
\hline
\textbf{\#} & \textbf{Application} & \textbf{Name} & \textbf{Summary} \\
\hline
2.1 & System & Reasonable precision and benchmarking across components & Component-level and end-to-end benchmarking and evaluations, where feasible. There is rarely ground truth GHG emissions data available. Additionally, experts do not always agree on how to generate a system model, LCI, or LCA for a given product. However, benchmarking is still useful across many aspects of assumptions and calculations to demonstrate rigor and demonstrate system improvements. \\
\hline
2.2 & System & Learning capability & Demonstrated improvement in accuracy and insight generation as more data is processed, via quality indicator improvements. \\
\hline
2.3 & Material & Uncertainty indication & Indication of confidence level, uncertainty ranges, or sensitivity for key datapoints. \\
\hline
\end{tabular}
\end{table}

\textbf{Criteria Details:}

\textbf{[2.1] Reasonable precision and benchmarking:} Carbon footprints and the underlying activity estimates and energy and processing components can be benchmarked against datasets. Full carbon footprints can be compared to results from published EPDs \cite{environdec}, licensed data such as ecoinvent \cite{ecoinvent}, and literature values such as The Carbon Catalogue \cite{meinrenken_carbon_2022}. For example, theoretical and practical energy minimums for industrial processes can be quantified using methodology proposed in \cite{cullen_framework_2025}. A potential target for full product footprints is median error and P90 error below specified thresholds when benchmarked against verified PCFs (where available). This is an area where publicly available datasets and leaderboards or competitions might be useful for ranking and calibrating different AI-assisted PCF systems. Care would be needed to ensure there is no cross-contamination of training data and evaluation data. 

\textbf{[2.2] Learning capability:} This can be shown via a narrative or examples for the results of actions taken on the match quality and uncertainty indicators; as more data is added, the quality should improve and the uncertainty should be reduced. There are many potential approaches to demonstrate a useful learning capability. It is related to uncertainty and will depend on how uncertainty is incorporated into and propagated through the system. Further work on the usefulness and best practices for decisive corporate action would be valuable to refine the approach. 

\textbf{[2.3] Display of uncertainty:} Modeled materials and products should make clear to the user which assumptions or nodes have the biggest uncertainties driving the overall uncertainty and sensitivity. This could be a simple indicator, or it could be numeric. PCFs contain a plethora of sources of uncertainty. A goal is to reduce these with a focus on where data can enable directionally correct comparative decisions. Lacking clear ground-truth data or sufficient sample sizes, it may be infeasible to achieve true statistically supported conclusions, which Henriksson et al. suggests is required for PCFs to be used to make decisions \cite{henriksson_2015}. Recent literature provides useful insights such as that the most relevant source of uncertainty for certain types of chemical production is in facility-level process specification, more than allocation method choices \cite{cullen_reducing_2024}. A key goal is to make a useful and actionable set of results rather than a mathematically rigorous theoretical model. 

Applications leveraging advanced AI capabilities should provide the indicators or written narratives shown in Table~\ref{tab:case2_transparency} at a system or material level, in addition to or as part of the metrics described above. 

\begin{table}[H]
\centering
\small
\caption{Case 2 transparency and traceability criteria}
\label{tab:case2_transparency}
\begin{tabular}{|p{.5cm}|p{2cm}|p{3.5cm}|p{8cm}|}
\hline
\textbf{\#} & \textbf{Application} & \textbf{Name} & \textbf{Summary} \\
\hline
2.4 & System & Key data sources & Documentation of major data sources and databases used. \\
\hline
2.5 & System & AI methodology & High-level explanation of the AI approaches used to generate insights beyond simple mapping. \\
\hline
2.6 & System & Summary methodology & High-level explanation of calculation approach. \\
\hline
2.7 & System & Limitations statement & Transparent communication of system limitations and appropriate levels. \\
\hline
2.8 & Material & Major assumptions & Recording of significant assumptions that drive results. \\
\hline
2.9 & Material & Hotspot identification & Clear indication of major emissions sources and improvement opportunities. \\
\hline
\end{tabular}
\end{table}

\textbf{Case 2 validation requirements}
Validation should include all of the same components as for Case 1, plus additional validation against the criteria described above for Case 2. This includes mainly system-level evaluation and validation, but the items such as user-displayed uncertainty indicators should also be applied to each individual product so that the user can better target data collection efforts. 

System-level review is an important aspect of scalable AI-CFs. The WBCSD PACT framework that compiles all the steps and processes for PACT-aligned PCFs \cite{wbcsd_pact} includes the concept of a dropdown of all steps and processes reviewed for PACT alignment during the PCF validation. This brings forth scalable concepts such as "3rd party assurance at PCF system level", boundary checks, and data input checks. If these checks can be performed on the PCF generation system using the artifacts described above, this type of approach may be more scalable than requiring each PCF to be 3rd party assured at the PCF level. 

The primary difference for bulk AI-CFs compared to traditional PCFs is that the criteria proposed here recommend that \textbf{system-level} assumptions, benchmarking, and evaluations suffice to achieve some criteria that traditionally would be reviewed on a line-by-line basis. By reviewing a sample of outputs systematically, reviewers and auditors should be able to generalize to the system. That is, not every single PCF must demonstrate independently that it meets all criteria. This will save effort in validation and review and therefore enable much more scaled PCF generation than would be feasible if every material in every PCF required full independent review of all assumptions. 

Non-exhaustive examples where system-level demonstrations of consistent application for example cases may be validated in lieu of PCF-level assumptions are as follows: 
\begin{itemize}
\item If the system follows boundaries provided as input methodology, then a validation for an AI-CF from that system does not need to separately confirm that the boundary for a particular methodology is followed for each PCF. 
\item If the system incorporates reference data or lookup tables provided as inputs, then a validation for an AI-CF from that system does not need to separately confirm that the specific reference data is used for each PCF. 
\item If the system is able to provide justification for key assumptions, not every key assumption needs to be reviewed. 
\item If the system uses reasonable formulas and assumptions or verified software for unit conversions, not every unit conversion in an AI-CF needs to be reviewed. 
\end{itemize}

These system-level reviews are a relatively novel approach to the sustainability field but are common in other disciplines e.g., software \cite{chen_system-level_2013,hand_validating_2020}, medicine \cite{allgaier_practical_2024, john_2023, straub_machine_2021}, with some generative AI specific approaches summarized in a recent literature review \cite{yu_measuring_2025}. Without system-level validation and verification, there is a risk that companies spend millions of dollars on repetitive third-party reviews for their AI-CFs before they can use them -- funds that would be better used for enacting sustainability initiatives. Once a system is reviewed, we recommend that reviewers and auditors still review the most important/critical outcomes, but no longer need to review inconsequential outcomes, similar to a GHG footprint audit.

\section{Discussion}
\label{sec:discussion}

Our findings indicate that the primary barrier to credible, at-scale AI-assisted carbon footprinting is not the technology itself, but the absence of a validation paradigm suited to its unique speed, scale, and algorithmic nature. A key finding of our research is that this emphasis on system-level validation with discrete evaluations for each major methodology pathway is useful for AI-assisted systems. Traditional, human-led carbon accounting involves a manageable number of discrete, explainable decisions, making manual, line-item verification feasible. In contrast, AI systems make thousands of algorithmic decisions, often within complex model architectures that are not easily audited on a case-by-case basis. Our consultations with industry experts confirmed that credibility and trust can be built without prohibitively expensive reviews. Therefore, our proposed criteria shift the focus from validating individual calculations to verifying the integrity, robustness, and logic of the entire system that produces them.

By implementing this case-dependent approach, organizations can deploy and validate AI-assisted PCF systems that are fit-for-purpose, cost-effective, and appropriate for their specific needs. 

Regardless of the use case, credible AI-assisted PCF systems require: 
\begin{itemize}
    \item \textbf{Transparency about limitations:} Clear communication of system capabilities and constraints.
    \item \textbf{User education:} Training users to understand outputs and their appropriate use.
    \item \textbf{Expert oversight:} Involvement of LCA experts in system development and validation.
\end{itemize}

Future advancements in AI-CF systems require continued research and development across several key areas. First, creating and sharing new, open-source \textbf{datasets} is crucial for evaluating system performance, including proxy mappings, data quality indicators, process assumptions, and end-to-end model accuracy. Second, the field would benefit from published \textbf{proxy frameworks} or hierarchies that provide a recommended and standardized approach for mapping to proxy data. Third, further work on \textbf{uncertainty quantification} is needed, including specific methodologies and case studies that demonstrate how uncertainty propagation can be a useful tool for corporate decision-making. Finally, the development of clear \textbf{standards compliance criteria} would help ensure that AI-CF systems can be reliably evaluated against specific, industry-recognized frameworks. Collectively, these efforts will support the evolution of more robust and effective AI-CF solutions.

\bibliography{references.bib}

\begin{thebibliography}{10}

\bibitem{adams_parameters_2025}
N~Adams and K~Allacker.
\newblock Parameter sensitivity and data uncertainty assessment of the cradle-to-gate environmental impact of state-of-the-art passive daytime radiative cooling materials.
\newblock {\em Environmental Sciences Europe}, 37(1):53, 2025.

\bibitem{neurips_tackling_2024}
Climate~Change AI.
\newblock Tackling {Climate} {Change} with {Machine} {Learning}, 2024.

\bibitem{neurips_tackling_2025}
Climate~Change AI.
\newblock Tackling {Climate} {Change} with {Machine} {Learning}, 2025.

\bibitem{ai4lca_2024}
{AI4LCA} {Workshop} 2024: {Accelerating} {Sustainability} {Through} {AI}-{Powered} {Life} {Cycle} {Assessment}, 2024.

\bibitem{aie_2025}
{AIE} {World}'s {Fair}, 2025.

\bibitem{allgaier_practical_2024}
Johannes Allgaier and Rüdiger Pryss.
\newblock Practical approaches in evaluating validation and biases of machine learning applied to mobile health studies.
\newblock {\em Communications Medicine}, 4(1):76, 2024.

\bibitem{balaji_emission_2025}
Bharathan Balaji, Fahimeh Ebrahimi, Nina~Gabrielle G~Domingo, Venkata Sai~Gargeya Vunnava, Abu-Zaher Faridee, Soma Ramalingam, Shikha Gupta, Anran Wang, Harsh Gupta, Domenic Belcastro, Kellen Axten, Jeremie Hakian, Jared Kramer, Aravind Srinivasan, and Qingshi Tu.
\newblock Emission factor recommendation for life cycle assessments with generative ai.
\newblock {\em Environmental Science \& Technology}, 59(18):9113--9122, 2025.

\bibitem{balaji_flamingo_2023}
Bharathan Balaji, Venkata Sai~Gargeya Vunnava, Nina Domingo, Shikhar Gupta, Harsh Gupta, Geoffrey Guest, and Aravind Srinivasan.
\newblock Flamingo: Environmental impact factor matching for life cycle assessment with zero-shot machine learning.
\newblock {\em ACM J. Comput. Sustain. Soc.}, 1(2):11, 2023.

\bibitem{basf_2025}
BASF.
\newblock Product {Carbon} {Footprint} of {Raw} {Materials}, 2025.

\bibitem{cullen_framework_2025}
Natanael Bolson, Luke Cullen, and Jonathan Cullen.
\newblock A robust framework for estimating theoretical minimum energy requirements for industrial processes.
\newblock {\em Energy}, 322:135411, 2025.

\bibitem{castle_entity_2025}
Steffen Castle, Julian~Moreno Schneider, Leonhard Hennig, and Georg Rehm.
\newblock Entity {Linking} using {LLMs} for {Automated} {Product} {Carbon} {Footprint} {Estimation}, 2025.

\bibitem{catena}
Catena-X.
\newblock Catena-x, 2025.

\bibitem{chen_system-level_2013}
Mingsong Chen, Xiaoke Qin, Heon-Mo Koo, and Prabhat Mishra.
\newblock {\em System-{Level} {Validation}: {High}-{Level} {Modeling} and {Directed} {Test} {Generation} {Techniques}}.
\newblock Springer, New York, NY, 2013.

\bibitem{cmr_2024}
CMR.
\newblock Why firms are not reporting their ghg emissions.
\newblock {\em California Management Review}, 2024.

\bibitem{epeat}
Global~Electronics Council.
\newblock Epeat criteria for solar products, 2025.

\bibitem{cullen_machine_2024}
Luke Cullen, Andrea Marinoni, and Jonathan Cullen.
\newblock Machine learning for gap-filling in greenhouse gas emissions databases.
\newblock {\em Journal of Industrial Ecology}, 28(4):636--647, 2024.

\bibitem{cullen_reducing_2024}
Luke Cullen, Fanran Meng, Rick Lupton, and Jonathan~M. Cullen.
\newblock Reducing uncertainties in greenhouse gas emissions from chemical production.
\newblock {\em Nature Chemical Engineering}, 1(4):311--322, 2024.

\bibitem{dumit_atlas_2024}
Andrew Dumit, Krishna Rao, Travis Kwee, Varsha Gopalakrishnan, Katherine Tsai, and Sangwon Suh.
\newblock {ATLAS}: {A} spend classification benchmark for estimating scope 3 carbon emissions.
\newblock In {\em Climate {Change} {AI}}. Climate Change AI, 2024.

\bibitem{ecoinvent}
ecoinvent database.

\bibitem{idf}
International~Dairy Federation.
\newblock Global carbon footprint standard for the dairy sector.

\bibitem{oecd}
Organization for Economic Development~(OECD).
\newblock Measuring carbon footprints of agri-food products: Eight building blocks, 2025.

\bibitem{wbcsd_pact}
The World Business~Council for Sustainable Development~(WBCSD).
\newblock Pact methodology.

\bibitem{wbcsd_validation}
The World Business~Council for Sustainable Development~(WBCSD).
\newblock Defining a practical and robust pcf validation approach, 2024.

\bibitem{ghgp_2025}
GHG Protocol Technical~Working Group.
\newblock Scope 3 presentation, 2025.

\bibitem{gomez-garza_barriers_2024}
Rodrigo Gómez-Garza, Leonor~Patricia Güereca, Alejandro Padilla-Rivera, and Alonso~Aguilar Ibarra.
\newblock Barriers and enablers of life cycle assessment in small and medium enterprises: a systematic review.
\newblock {\em Environment, Development and Sustainability}, 2024.

\bibitem{hand_validating_2020}
David~J. Hand and Shakeel Khan.
\newblock Validating and verifying ai systems.
\newblock {\em Patterns}, 1(3):100037, 2020.

\bibitem{henriksson_2015}
Patrik J.~G. Henriksson, Reinout Heijungs, Huy~Manh Dao, Long~Tran Phan, Geert~R. de~Snoo, and Jeroen~B. Guinée.
\newblock Product carbon footprints and their uncertainties in comparative decision contexts.
\newblock {\em PLoS ONE}, 10(3):e0121221, 2015.

\bibitem{environdec}
EPD International.
\newblock Epd library.

\bibitem{isie_2025}
Applications of {AI} in streamlining the lca modeling data, 2025.

\bibitem{iso_14044}
ISO.
\newblock {ISO} 14044:2006 - environmental management — life cycle assessment — requirements and guidelines, 2006.

\bibitem{iso_14067}
ISO.
\newblock {ISO} 14067:2018 - {Greenhouse} gases — {Carbon} footprint of products, 2018.

\bibitem{john_2023}
S.~H. John.
\newblock Validation in the age of machine learning: A framework for describing validation with examples in transcranial magnetic stimulation and deep brain stimulation.
\newblock {\em Intelligence-Based Medicine}, 7:100090, 2023.

\bibitem{konradsen_same_2024}
Freja Konradsen, Kristine Sofie~Holse Hansen, Agneta Ghose, and Massimo Pizzol.
\newblock Same product, different score: how methodological differences affect {EPD} results.
\newblock {\em The International Journal of Life Cycle Assessment}, 29(2):291--307, 2024.

\bibitem{kpmg_2025}
KPMG.
\newblock Validating {AI} models, 2025.

\bibitem{kuczenski_prototypes_2021}
Brandon Kuczenski, Chris Mutel, Michael Srocka, Kelly Scanlon, and Wesley Ingwersen.
\newblock Prototypes for automating product system model assembly.
\newblock {\em The International Journal of Life Cycle Assessment}, 26(3):483--496, 2021.

\bibitem{lcic_workshop_2024}
Workshop 04: {Harnessing} the {Power} of {AI} for {Enhanced} {Life} {Cycle} {Assessment} and {Ecodesign}, 2024.

\bibitem{li_pcf-rwkv_2025}
Zhen Li, Peihao Tang, Xuanlin Wang, Xueping Liu, and Peng Mou.
\newblock {PCF}-{RWKV}: {Large} {Language} {Model} for {Product} {Carbon} {Footprint} {Estimation}.
\newblock {\em Sustainability}, 17(3):1321, 2025.

\bibitem{martinez-ramon_frameworks_2024}
Nicolás Martínez-Ramón, Fernando Calvo-Rodríguez, Diego Iribarren, and Javier Dufour.
\newblock Frameworks for the application of machine learning in life cycle assessment for process modeling.
\newblock {\em Cleaner Environmental Systems}, 14:100221, 2024.

\bibitem{meinrenken_carbon_2022}
Christoph~J. Meinrenken, Daniel Chen, Ricardo~A. Esparza, Venkat Iyer, Sally~P. Paridis, Aruna Prasad, and Erika Whillas.
\newblock The {Carbon} {Catalogue}, carbon footprints of 866 commercial products from 8 industry sectors and 5 continents.
\newblock {\em Scientific Data}, 9(1):87, 2022.

\bibitem{mendoza_beltran_quantified_2018}
Angelica Mendoza~Beltran, Valentina Prado, David Font~Vivanco, Patrik J.~G. Henriksson, Jeroen~B. Guinée, and Reinout Heijungs.
\newblock Quantified {Uncertainties} in {Comparative} {Life} {Cycle} {Assessment}: {What} {Can} {Be} {Concluded}?
\newblock {\em Environmental Science \& Technology}, 52(4):2152--2161, 2018.

\bibitem{neurips_2024}
Neurips 2024, 2024.

\bibitem{ghgp_product}
GHG Protocol.
\newblock Product standard.

\bibitem{ghgp_uncertainty}
GHG Protocol.
\newblock Ghg protocol guidance on uncertainty assessment in ghg inventories and calculating statistical parameter uncertainty, 2023.

\bibitem{qiao_stochastic_2025}
Yaning Qiao, Xia Wen, Shirui Liu, Songtao Lv, and Liang He.
\newblock Stochastic analysis for comparing life cycle carbon emissions of hot and cold mix asphalt pavement systems.
\newblock {\em Resources, Conservation and Recycling}, 212:107881, 2025.

\bibitem{rivalin_2024}
Lisa Rivalin, Lingyun Yi, Megan Diefenbach, Alex Bruefach, Frances Amatruda, and Tobias Tiecke.
\newblock Estimating embodied carbon in data center hardware, down to the individual screws, 2024.

\bibitem{rivian_2025}
Rivian.
\newblock R1 {Gen} 2 {Carbon} {Footprint} {Reports}, 2025.

\bibitem{straub_machine_2021}
Jeremy Straub.
\newblock Machine learning performance validation and training using a 'perfect' expert system.
\newblock {\em MethodsX}, 8:101477, 2021.

\bibitem{sturm_systematic_2025}
Johannes H.~L. Sturm, Sebastian Gehrke, and Christoph Herrmann.
\newblock A {Systematic} {Approach} for the {Reliable} and {Automated} {Selection} of {Life} {Cycle} {Assessment} {Data} {Sets} {Exemplified} by the {Automotive} {Industry}.
\newblock In {\em Circularity {Days} 2024}, pages 393--406. Springer Fachmedien, 2025.

\bibitem{tascione_comparative_2024}
Valentino Tascione, Alberto Simboli, Raffella Taddeo, Michele Del~Grosso, and Andrea Raggi.
\newblock A comparative analysis of recent life cycle assessment guidelines and frameworks: {Methodological} evidence from the packaging industry.
\newblock {\em Environmental Impact Assessment Review}, 108:107590, 2024.

\bibitem{trek_2025}
Trek.
\newblock Trek {Bicycle} {Announces} {Industry}-{First} {Shift} to {Low}-{Emission} {Aluminum}, 2025.

\bibitem{unilever_2022}
Unilever.
\newblock Using innovation to lower the footprint of our products, 2022.

\bibitem{wakelin_how_2021}
Nicole Wakelin.
\newblock How {Many} {Parts} are in a {Car}?, 2021.

\bibitem{wang_2025}
Anran Wang, Zaid Thanawala, Harsh Gupta, Jeremie Hakian, Jared Kramer, Kommy Weldemariam, and Bharathan Balaji.
\newblock Palimpsest: Bill of materials prediction - a case study with solid state drives.
\newblock In {\em ICLR 2025 Climate Change AI Workshop}, 2025.

\bibitem{wasserman_ai-powered_2024}
Molly Wasserman.
\newblock {AI}-{Powered} {Life} {Cycle} {Assessments} ({LCAs}), 2024.

\bibitem{useeio_2025}
Ben Young and Wesley Ingwersen.
\newblock {USEEIO} v2.5 {Models}, 2025.

\bibitem{yu_measuring_2025}
Liang Yu, Emil Alégroth, Panagiota Chatzipetrou, and Tony Gorschek.
\newblock Measuring the quality of generative {AI} systems: {Mapping} metrics to quality characteristics — {Snowballing} literature review.
\newblock {\em Information and Software Technology}, 186:107802, 2025.

\bibitem{yu_sensitivity_2017}
Xi~Yu, Haiqing Zhang, Hongping Shu, Weidong Zhao, Tao Yan, Yonghong Liu, and Xie Wang.
\newblock A robust eco-design approach based on new sensitivity coefficients by considering the uncertainty of lci, 2017.

\bibitem{zhang_towards_2025}
Zhihan Zhang, Alexander Metzger, Yuxuan Mei, Felix Hähnlein, Zachary Englhardt, Tingyu Cheng, Gregory~D. Abowd, Shwetak Patel, Adriana Schulz, and Vikram Iyer.
\newblock Towards {Autonomous} {Sustainability} {Assessment} via {Multimodal} {AI} {Agents}, 2025.

\bibitem{zhao_data-centric_2025}
Bu~Zhao, Jitong Jiang, Ming Xu, and Qingshi Tu.
\newblock A data-centric investigation on the challenges of machine learning methods for bridging life cycle inventory data gaps.
\newblock {\em Journal of Industrial Ecology}, 29(3):955--966, 2025.

\end{thebibliography}
\end{document}